\documentclass[12pt]{article}
\usepackage{latexsym}
\usepackage{epsfig,amssymb,euscript}
\usepackage{amsmath}

\def\be{\begin{equation}}
\def\ee{\end{equation}}
\def\bea{\begin{eqnarray}}
\def\eea{\end{eqnarray}}

\input epsf
\begin{document}
\font\cmss=cmss10 \font\cmsss=cmss10 at 7pt
\hfill SISSA-89/2004/EP
\par\hfill LPTHE-04-31
\par\hfill CPHT-RR063.1104
\par\hfill hep-th/0411249\\
\vspace{10pt}
\begin{center}
{\LARGE \textbf{New checks and subtleties for \vskip 15pt AdS/CFT and $a$-maximization}}
\end{center}

\vspace{10pt}

\begin{center}
{\Large M. Bertolini$\,^a$, F. Bigazzi$\,^{b,c}$, A. L. Cotrone$\,^{d}$} \\ 
\end{center}
\vskip 10pt
\begin{center}
\textit{a} SISSA/ISAS and INFN, Via Beirut 2; I-34014 Trieste, Italy.\\
\textit{b} LPTHE, Universit\'es Paris VI and VII, 4 place Jussieu; 75005, Paris, France.\\
\textit{c}  INFN, Piazza dei Caprettari, 70; I-00186  Roma, Italy.\\
\textit{d} CPHT,  \'Ecole Polytechnique, 48 Route de Saclay; F-91128 Palaiseau Cedex, France.
\vskip 8pt
{\small\tt bertmat@sissa.it, bigazzi@lpthe.jussieu.fr, Cotrone@cpht.polytechnique.fr}
\end{center}

\vspace{15pt}

\begin{center}
\textbf{Abstract}
\end{center}

{\small \noindent
We provide a cross-check of AdS/CFT and $a$-charge maximization for a four dimensional 
${\cal N}=1$ SCFT with irrational $R$-charges. The gauge theory is the low energy effective 
theory of $N$ D3-branes at the tip of
the complex cone over the first del Pezzo surface. By carefully
taking into account the subtle issue of flavor symmetry breaking at the fixed point, we show, 
using $a$-maximization, that this theory has in
fact irrational central charge and $R$-charges. Our results perfectly match with those inherited from the 
recently discovered supergravity dual background. Along analogous lines, 
we make novel predictions for the still unknown AdS dual of the quiver 
theory for the second del Pezzo surface. This should flow to
a SCFT with irrational charges, too. All of our results differ from previous 
findings in the literature and outline interesting subtleties in $a$-maximization and 
AdS/CFT techniques overlooked in the past.}
\vfill

\newpage

\section{Overview}

Placing D3-branes at conical singularities is a useful tool to obtain
four dimensional SCFT's with ${\cal N}<4$ supersymmetries. If the cone is a
Calabi-Yau threefold, the low energy theory on the branes is
an ${\cal N}=1$ SCFT. The AdS/CFT correspondence \cite{Maldacena}
relates the latter to type IIB string theory on $AdS_5 \times X^5$,
where $X^5$, the base of the cone, is a Sasaki-Einstein (SE) manifold
\cite{conic,kw}.
A deeply explored example is for $X_5=T^{1,1}$ \cite{kw} which has
topology $S^2\times S^3$. This is a ``regular'' SE manifold, as it
arises as a free $U(1)$ fibration over a  K\"ahler-Einstein space
($S^2\times S^2$). The class of regular SE manifolds is completely
known and includes only few other cases. The less trivial ones (their
metrics being unknown) arise when the fibration is over a del Pezzo
surface $dP_k$, obtained as the blow up of $\mathcal{CP}^2$ at
$k=3,4,5,6,7,8$ points.

Recently a remarkable result was obtained in \cite{gmswSE}, where an infinite number of 
new smooth SE manifolds and their metrics were found. These manifolds were dubbed
$Y^{p,q}$, where $p,q$ are co-prime integer labels with $q<p$,  \cite{gmswSE} 
(we refer to this paper for any detail on the geometry). Their
topology is $S^2 \times S^3$ and they have $SU(2)\times U(1)\times
U(1)$ isometry group. These manifolds are either ``quasi-regular'' (the $U(1)$
fibration is over a space with orbifold singularities) or
``irregular'' (the Killing vector has non compact orbits). The
quasi-regular (resp. irregular) spaces have volumes which are rational
(resp. irrational) multipliers of the volume of a unit round $S^5$
\be
\label{volume}
V(Y^{p,q})=\frac{q^2\,\left(2p+\sqrt{4p^2-3q^2}\right)}{3p^2\,
\left(3q^2-2p^2+p\,\sqrt{4p^2-3q^2} \right)}\,\pi^3\ .
\ee
AdS/CFT relates $V(X_5)$ to the $a$-central charge\footnote{Here $TrR$ and 
$TrR^3$ are the linear and cubic 't Hooft anomalies for the exact $R$-symmetry. Notice
that for generic CFT's $a$ differs from the central charge $c =
(1/32) (9 Tr R^3 - 5 Tr R)$. However, for large $N$ theories having
a holographic dual, $Tr R=0$ and hence $a=c$ \cite{HS}.}
\be
a = \frac{3}{32} \left(3 Tr R^3 - Tr R \right)
\label{ace}
\ee
of the dual IR SCFT via the holographic relation \cite{Maldacena}
\be
\label{volumes}
V(X^5) =  \frac{N^2}{4\,a(X^5)}\,\pi^3 \ ,
\ee
where $N$ is the number of D3-branes at the tip of the cone.
Thus, for the irregular SE manifolds of \cite{gmswSE} AdS/CFT predicts that the dual field theories 
have irrational $a$-charge, and hence irrational $R$-charges at the IR fixed point.

In this paper we test this prediction for a particular
case of those studied in \cite{gmswSE}, $Y^{2,1}$, for which, due to recent 
geometric findings\footnote{We thank the authors, 
D. Martelli and J. Sparks, for having made us
aware of their results before their paper was published.} \cite{MS}, the dual
field theory is known. Indeed $Y^{2,1}$ is shown to be the horizon of the complex 
cone over the first del Pezzo surface $dP_1$. The dual
field theory was built for this case years ago (see for instance 
\cite{hanany}): it is a quiver theory with gauge group $SU(N)^4$, 
bi-fundamental chiral fields and a superpotential at the IR fixed point. The
supergravity prediction for $a$, using Eqs.~(\ref{volume}), (\ref{volumes}),
is
\be
a(Y^{2,1})= N^2(-46+13\sqrt{13})\ .
\label{gpred}
\ee 
Another prediction refers to the $R$-charges of baryons in the dual CFT. In 
fact, D3-branes
wrapped on supersymmetric three-cycles $\Sigma_A$ in the dual geometry correspond to baryons whose 
$R$-charge is proportional to the volumes of such
three-cycles via the holographic relation \cite{W,BHK}
\be
\label{barvol0}
R(B_A) ~=~ \frac{\pi}{3} \; \frac{V(\Sigma_A)}{V(X^5)} ~N \ .
\ee
It is shown in \cite{MS} that each SE manifold $Y^{p,q}$ admits 
two such three-cycles whose volumes, for the case 
we are interested in, are
\be
\label{3cycles}
V(\Sigma_1) = \frac{1}{36} \left(7 + \sqrt{13}\right)\, \pi^2~~~,~~~
V(\Sigma_2) = \frac{1}{108} \left(31 + 7 \sqrt{13}\right)\, \pi^2~~~.
\ee
Hence, using (\ref{barvol0}), we should expect irrational baryon 
charges in the dual CFT.

Interestingly, for the case at hand the literature offers a
purely field theoretical calculation of $a$, \cite{IW}. The result was found
using the ``$a$-maximization prescription'' proposed by Intriligator 
and Wecht (IW) in \cite{IW}. This is a
remarkable prescription that allows one to determine the exact
$R$-charges of a CFT in those cases where these cannot be completely fixed by
(super) symmetries or (ABJ) anomaly-vanishing arguments. This happens when the 
SCFT admits some global symmetry group containing $U(1)$ factors commuting 
with the non-abelian flavor symmetries and mixing with $U(1)_R$ \footnote{We will 
refer to $U(1)_R$ as one of the possible $R$-symmetries, dubbed $R_0$ in \cite{IW}.}
(this is the case 
for the theory for $dP_1$ \footnote{For regular SE manifolds, instead, the symmetries 
are enough to fix $R$.}). 
The $a$-maximization prescription states 
that among the possible choices, the (a priori arbitrary) combination of $U(1)$-charges corresponding to 
the exact $R$-symmetry is the one which locally maximizes $a$. We will point out
that particular care is needed in including, among these $U(1)$'s, the ones eventually coming from the 
breaking of non abelian flavor symmetries in the IR. We 
refer to \cite{IW,IW2,KPS,kutasov,IW3,KS} for more details on $a$-maximization.

The result found in \cite{IW} for the quiver theory for the first del Pezzo surface 
is that $a =(27/32)N^2$.
Similarly, the $R$-charges for the baryons were computed, again predicting rational
values \cite{IW2,HW,Han}. At that time this did not produce any puzzle, since 
AdS computations where based on the (wrong, as the results in 
\cite{MS} have now explicitly shown) assumption that the 
geometric results would not have been affected by the manifold being an 
irregular SE manifold.\footnote{In the regular cases, one can easily prove 
that the volumes are rational.} However, these field theory findings contradict 
the more recent geometrical predictions 
(\ref{gpred}) and (\ref{3cycles}), opening a possible puzzle for AdS/CFT.

In Section 2 we solve the puzzle. Carefully taking into account the flavor symmetry breaking
pattern driven by the superpotential (a subtlety which was 
overlooked in literature) we calculate the $a$-charge and the $R$-charges by using
$a$-maximization and find perfect agreement with the results of \cite{MS}, 
Eqs.~(\ref{gpred}) and (\ref{3cycles}). This shows that the exact $R$-charges 
for the theory on $N$ D3-branes at the tip of the complex cone over $dP_1$ are indeed irrational. 
To our knowledge this is the 
first example of a CFT with irrational $R$-charges having a known
holographic dual. Our calculation provides a cross-check of AdS/CFT and $a$-maximization in 
a (till now) unexplored realm. It also points into refining part of the AdS/CFT techniques for the case
of non-regular SE manifolds, as suggested in \cite{MS}.

In Section 3 we reexamine the SCFT living 
on $N$ D3-branes on the complex cone over the second del Pezzo surface, 
$dP_2$. This is not  K\"ahler-Einstein, just as $dP_1$, and so we expect this case to be 
subtle, too. The field theory \cite{hanany} has $SU(N)^5$ gauge group and bi-fundamental 
chiral matter content. Moreover, 
there is a superpotential, which crucially breaks all the non-abelian flavor 
symmetries to abelian subgroups. Assuming that the theory flows to an interacting fixed point 
and using $a$-maximization, we show
that the SCFT at hand has irrational $R$-charges. This is again in contradiction 
with previous findings \cite{IW2,HW,Han}. Our results also provide a purely field theoretical prediction 
on details of the dual (and so far unknown) geometry.

\vskip 5pt
The two main results of this paper can then be summarized as follows:
\begin{itemize}
\item{From a field theoretical point of view, our findings outline the 
following simple, but crucial, issue. As remarked in \cite{IW}, the 
non-abelian flavor symmetries do not contribute to the exact $R$. However, 
care is needed in identifying these symmetries {\it in the IR}. A superpotential at the
IR fixed point, for example, can break some non abelian flavor
symmetries to give extra abelian factors. The latter {\it can and do mix} with
the other $U(1)$'s and contribute in determining the exact $R$-symmetry. 
}
\item{From the AdS/CFT point of view, we provide a non-trivial check for 
the duality discussed above and a solution of the 
related puzzle.
Moreover, we give a new argument in support of this duality, showing that the symmetries of the supergravity background
exactly match the ones in field theory.
Finally, our results explicitly 
stress, as remarked in \cite{MS}, the peculiarities of AdS/CFT techniques in cases with irregular SE factors.}
\end{itemize}

Now that the AdS/CFT issues for the theory on $dP_1$ 
have been clarified, it would be very interesting to add fractional branes to this 
system \cite{FHHW}, as it was done for the conifold \cite{KT,KS2}. This would break conformal 
invariance and could reveal interesting 
dynamics. Work is in progress in this direction \cite{BBC}. Another interesting 
development would be to study the KK spectrum of type IIB on 
$AdS_5 \times Y^{2,1}$, similarly to what has been done for  
$T^{1,1}$ \cite{fer}. This would help in determining the exact bulk field/CFT operator map.
Finally, the SCFT duals of general $Y^{p,q}$ have to be uncovered.

\section{The quiver gauge theory for the first del Pezzo}
\begin{figure}[ht]
\begin{center}
{\includegraphics{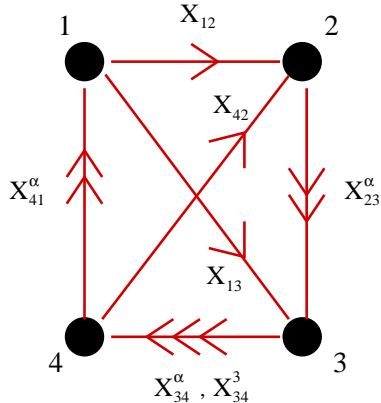}}
\caption{\small The quiver associated to the first del Pezzo 
surface. Each dot represents a $SU(N)$ factor while each arrow
represents a bi-fundamental chiral multiplet.}
\label{reg1}
\end{center}
\end{figure}

The quiver gauge theory for $dP_1$ was constructed in \cite{hanany,hanany2} and is a four dimensional 
${\cal N}=1$ SCFT with gauge group $SU(N)^4$ and bi-fundamental matter. The corresponding 
quiver diagram is depicted in Figure \ref{reg1}. 
Beside the four gauge multiplets, there are also ten chiral multiplets
$X_{ij}$. The notation is such that the chiral superfield $X_{ij}$
transforms in the $(N_i,\bar N_j)$. The flavor symmetry group of the theory without superpotential 
\cite{IW} is $SU(3)\times SU(2)\times SU(2)\times U(1)\times U(1)$.

At the IR fixed point it is believed that the theory has also a superpotential given by
\be
\label{super}
W= \mbox{Tr}\;\left[\epsilon_{\alpha\beta}\,X^\alpha_{34} X^\beta_{41}X_{13}-\epsilon_{\alpha\beta}\, 
X^\alpha_{34} X_{42}X^\beta_{23} + \epsilon_{\alpha\beta}\,X^3_{34} X^\alpha_{41}X_{12}X^\beta_{23}\right] \ ,
\ee
where $\alpha,\beta=1,2$. This explicitly breaks the $SU(3)$ factor
(associated to the 3 bi-fundamental fields  $X^I_{34},\,I=1,2,3$)
of the flavor symmetry group to $SU(2)\times U(1)$, as can be seen by
the different r\^ole  played by the doublet $X^\alpha_{34}$ and the
singlet $X^3_{34}$ in the above expression. Similarly, only a single 
diagonal $SU(2)$ survives and a diagonal $U(1)$ of the 
two abelian factors. Therefore, the residual non-$R$ 
symmetry group at the IR fixed point is $SU(2)\times U(1)\times
U(1)$. Hence, according to what summarized previously, there 
are two abelian factors which are expected to mix with $U(1)_R$ 
and give the actual exact $R$-symmetry of the IR SCFT. 

Note that, as outlined in \cite{IW}, the theory at hand could be
potentially problematic. For two of the nodes in the quiver ($3,4$ in
our notations) the number of fundamental (and anti-fundamental) fields
is $N_f=3N_c=3N$. This means that both factors lye at the boundary of
the Seiberg conformal window $(3/2)N < N_f < 3N$ and in particular their
one-loop beta function vanishes in the UV. This should imply that both
factors are IR free instead of being interacting. We will assume that the
whole theory flows to an IR interacting fixed point. 
We will also assume that the gauge couplings in the IR are sufficiently 
small to avoid singularities in the $\beta$-functions. The agreement we
will find with AdS dual results strengthens these assumptions.

\subsection{Calculating the central charge via $a$-maximization}

Let us discuss the conditions imposed on the $R$-charges by the fact that the theory 
flows to an IR SCFT with superpotential. We adopt here the same logic as in \cite{kutasov}.

Generically, the $\beta$-functions for each quiver factor are proportional to
\be
\beta \sim 3 \, T(G) - \sum_A T(r_A) \left(1 - \gamma_A \right)\ ,
\ee
where the scaling dimension of a chiral field $X_A$ is $\Delta_A = 1 + \gamma_A/2$, and it is related to 
the $R$-symmetry charge by $\Delta_A = \frac 32\, R_A$. $T(G)$ is the Casimir of the adjoint, and $T(r_A)$ the 
Casimir of the representation $r_A$ under which the field $X_A$ transforms. For $G=SU(N)$, $T(G)=N$ 
and $T(\rm{fund})=1/2$. At the IR interacting fixed point 
the $\beta$-functions of the four gauge groups should vanish and this gives the conditions (with 
obvious notations, we suppress redundant $SU(2)$ indexes)\footnote{These conditions are equivalent to the 
ABJ anomaly vanishing conditions.}
\bea
\label{betas1}
{\hat\beta}_{1\star}&\equiv& N + \frac{1}{2}N (R_{12} - 1) + \frac{1}{2}N (R_{13} - 1) +
N(R_{14} - 1) = 0\ , \\
\label{betas2}
{\hat\beta}_{2\star}&\equiv&  N + \frac{1}{2}N (R_{12} - 1) + N (R_{23} - 1) 
+\frac{1}{2}N(R_{42} - 1) = 0\ ,  \\
\label{betas3}
{\hat\beta}_{3\star}&\equiv&  N + \frac{1}{2}N (R_{13} - 1) + N (R_{23} - 1) 
+N(R^{(1)}_{34} - 1)+\frac{1}{2}N (R^{(3)}_{34} - 1)  = 0\ , ~~~ \\
\label{betas4}
{\hat\beta}_{4\star}&\equiv&  N + N (R_{14} - 1) + \frac{1}{2}N (R_{24} - 1) 
+N(R^{(1)}_{34} - 1)+\frac{1}{2}N (R^{(3)}_{34} - 1)  = 0\ , ~~~
\eea
where $\star$ means ``at the fixed point''. Following IW's notations, we moved to the double 
index notation adapted to this quiver theory, $A \rightarrow ij$, and we used the fact that 
$1-\gamma_{ij} = 3 \left(1- R_{ij} \right)$ for chiral fields. Moreover, constrained by the $
SU(3)\rightarrow SU(2)$ symmetry breaking dictated by the superpotential (\ref{super}) we are 
taking the still 
unknown charges $R^{(3)}_{34}$ and $R^{(1)}_{34}=R^{(2)}_{34}$ as different ones.\footnote{Our choices 
of the $R$-charges could be rendered more clear by redefining the fields in such a way that $W$ 
becomes manifestly invariant under the diagonal $SU(2)$. One can show that this would produce the same results 
we find here.} 
This point was overlooked in the literature and, as we are going to show, plays a crucial r\^ole.

Since, in order to compare our results with the string/supergravity predictions, we are interested in the 
large $N$ limit,  we will ignore all terms of order $1/N^k$ with $k > 0$. It is 
easy to verify that in this case 
\be
Tr \, R \equiv \sum_{ferm} r(ferm) = N \sum_{i=1}^4{\hat\beta}_{i\star}=0 \ ,
\ee
as expected since the theory has a holographic dual. 
We see that the above condition does not give a new constraint with respect to  
Eqs.~(\ref{betas1})-(\ref{betas4}).

A genuinely new constraint on the $R$-charges comes from the superpotential. Indeed, we have to impose that 
the superpotential has $R$-charge equal to 2 at the IR fixed point. This gives the conditions
\bea
\label{cW1}
&&R^{(1)}_{34}+R_{41}+R_{13} = 2\ , \\
&&R^{(1)}_{34}+R_{23}+R_{42} = 2\ , \\
\label{cW3}
&&R^{(3)}_{34}+R_{12}+R_{41}+R_{23} = 2\ .
\eea
The conditions (\ref{betas1})-(\ref{betas4}) and (\ref{cW1})-(\ref{cW3}) leave us with only {\it two}
unknowns to be determined by $a$-maximization. These precisely correspond to the two $U(1)$ factors 
of the global symmetry group which mix with $U(1)_R$ 
to give the exact $R$-symmetry in the IR. We have
\bea
\label{sist}
R_{13}=R_{42}=R^{(3)}_{34} \equiv x &,&  R_{14}=R_{23}\equiv y \ , \nonumber \\
R^{(1)}_{34}=R^{(2)}_{34}=2-x-y &,&  R_{12}=2-x-2y \ .
\eea
Using the above values, we find for the``trial''(i.e. the one before maximization), 
$a$-charge $a_t$ (\ref{ace}) the $(x,y)$-dependent expression
\be
\label{atrial}
a_t(x,y) = \frac{9 N^2}{32}\left[ 4 + 4 (y-1)^3 + 3 (x-1)^3 + (1 -x -2y)^3 + 2 (1-x-y)^3\right]\ .
\ee
Extremizing with respect to $x$ and $y$ we find that the maximum for $a_t$ is at
\be
\label{sol2}
x_{m} = -3 + \sqrt{13} ~~,~~ y_{m}  = \frac{4}{3} \left(4 - \sqrt{13} \right)\ ,
\ee
giving an {\it irrational} central charge
\be
\label{cIR}
c=a= a_t(max) \,=\, N^2 \, \left(-46 + 13 \sqrt{13} \right)\ .
\ee

It is a simple exercise to show that without implementing the flavor symmetry breaking pattern discussed 
above, hence taking $R^{1}_{34}=R^{2}_{34}=R^{3}_{34}$, the conditions on the vanishing $\beta$-functions 
and on the $R$-charge of the superpotential leave only one $R$-charge to be determined. Using $a$-maximization 
the value one would find is rational, $a=27N^2/32$. The same happens for the 
$R$-charges.

A technical remark. As pointed out in \cite{IW}, the $a$-charge
``knows'' about the {\it unbroken} non-abelian flavor symmetries, and
the fact that they do not contribute to the exact $R$-charge. So, for
example, if one considers the theory above  without the superpotential
(\ref{super}), there are only two $U(1)$'s that can mix with
$U(1)_R$. However, if one performs the $a$-maximization keeping
different $R$-charges for all the ten chiral multiplets, one obtains, after
imposing the vanishing of the $\beta$-functions, a six-parameter
dependent $a_{t}$. Nevertheless, its actual maximum $a_t(max)$ and the
$R$-charges turn out to be the correct ones calculated in \cite{IW} for the 
case $W=0$. The same conclusions applies to the $W\neq0$ theory above.

Therefore, the final lesson is that whenever it is hard to understand which 
symmetries are preserved in the IR, one can drop this complication and just keep all the $R$-charges 
different, being guaranteed  that anyway $a_t(max)$ will be the correct one. As our calculation shows, 
the important point is not to over-constrain the $R$-charges, as this obviously affects the determination 
of the correct maximum of $a_t$. 

\subsection{The $R$-charges}
Plugging the solution (\ref{sol2}) into the relations (\ref{sist}) we get the $R$-charges for 
the bi-fundamental fields of the theory. These are reported in Table 1.
\begin{table} [ht]
\label{tabar}
\begin{center}
\begin{tabular}{|c|c|}
\hline        
  $X_{ij}$          & $R_{ij}$ \\ 
\hline  
  $X_{12}$          & $\frac 13 (-17 + 5 \sqrt{13})$ \\  
  $X^\alpha_{23},\, X^\alpha_{41}$   & $\frac 43 (4 - \sqrt{13})$ \\  
  $X^\alpha_{34}$  & $\frac 13 (-1 + \sqrt{13}$) \\  
  $X^3_{34},\,X_{13},\,X_{42} $       & $- 3 + \sqrt{13}$ \\  
  \hline \end{tabular}
\caption{\small The exact $R$-charges of the bi-fundamental chiral fields for the $dP_1$ dual.}
\end{center}
\end{table} 

It can be checked that with the above assignment the theory does not violate the unitarity bound. 
For the gauge invariant mesons not entering the superpotential, 
$X^3_{34} X^\beta_{41}X_{13}$, $X^3_{34} X_{42}X^\beta_{23}$ and 
$X^\gamma_{34} X^\alpha_{41}X_{12}X^\beta_{23}$, this is immediately verifiable, being 
their conformal dimension $\Delta>1$.

The baryons are color singlets defined from the corresponding bi-fundamental chiral fields as
\be
\label{bar}
B_{ij} = det_{N\times N} \left( X_{ij} \right)\ . 
\ee
Hence they do not give any problem to the unitarity bound: their $R$-charge goes like $N$, since 
$R(B_{ij}) = N \,R_{ij}$.

\subsection{A check for AdS/CFT}

In \cite{MS} it is argued that the field theory analyzed above should be dual to type IIB 
string theory on $AdS_5 \times Y^{2,1}$. As we are going to show, our results indeed confirm this prediction 
and thus provide the first example of gauge/gravity duality for SCFT's with irrational $R$ charges. 

The first important check concerns the central charge. The irrational $a$-charge we have obtained, 
Eq.~(\ref{cIR}), is in perfect agreement with the supergravity prediction \cite{MS}, Eq.~(\ref{gpred}).
\footnote{An analogous striking test of AdS/CFT was obtained in the 
regular case of the conifold \cite{gubser}. In that case, however, no $a$-maximization is required to fix the $R$-charges.}

Another matching is provided by baryon $R$-charges: 
using the holographic relations (\ref{volumes}) 
and (\ref{barvol0}), we get for the dual volumes of the corresponding three-cycles   
\be
V(\Sigma_A)  ~=~ \frac{3}{4} N \pi^2 \, \frac{1}{a} \, R(B_A) =  \left\{\begin{array}{cc} 
\pi^2 \,\left(7 + \sqrt{13} \right)/36 &  \hskip -25pt A = (12)\\ 
\pi^2 \,\left(31 + 7 \sqrt{13} \right)/108 & ~~~~~~~  A = (34), (13), (42)
\end{array} \right.
\label{voldual}
\ee
which is nothing but 
Eq.~(\ref{3cycles}), that is what computed for the dual geometry in \cite{MS}.

Finally, 
let us add another new piece of evidence in support of the duality, 
showing that
also the global symmetries match. The field theory global
symmetries are the exact $R$-symmetry just computed and the global
$SU(2)\times U(1)\times U(1)$.  The $Y^{2,1}$ metric only admits a
$SU(2)\times U(1) \times U(1)$ isometry \cite{gmswSE}. However, the
whole supergravity background contains also a non-trivial flux of the
self dual RR five-form field strength $F_5=dC_4$. The KK reduction of
$C_4$ on the unique independent three-cycle of $Y^{2,1}$ (whose topology
is in fact $S^2 \times S^3$) provides an additional $U(1)_B$ baryonic
symmetry. Hence the continuous global symmetry group of the dual field
theory is expected to be  \be SU(2) \times U(1) \times U(1)\times
U(1)_B\ .  \ee This precisely agrees with the symmetries of the field
theory above. One of the three $U(1)$'s in the latter is thus mapped
to $U(1)_B$.
Let us notice that the field theory at hand is a chiral quiver. Then
the baryonic $U(1)_B$ do mix with $U(1)_R$ to give the exact
$R$-symmetry, as remarked in \cite{IW2}. This is different to what
happens in non-chiral theories where baryonic symmetries commute with
$R$ \cite{IW}.\footnote{We are grateful to Ken Intriligator for
a crucial observation on this point.}

\section{The SCFT for the second del Pezzo surface and a prediction for its AdS dual}

The second del Pezzo surface $dP_2$ shares with $dP_1$ the property of not admitting a K\"ahler-Einstein 
metric \cite{TY}. The base of the complex cone over $dP_2$ is thus a non-regular SE manifold and its 
volume is not guaranteed to be rational \cite{BH}. Therefore, one could not exclude to get a dual SCFT 
with irrational $R$-charges by considering D3-branes at the tip of that cone. The dual gauge theory was 
worked out in \cite{hanany,hanany2} and we can now repeat the same rationale we pursued for the case of $dP_1$. 
In this case neither we know the AdS dual, nor a metric for $dP_2$. So, our results 
should furnish novel predictions on the dual geometry.

\begin{figure}[ht]
\begin{center}
{\includegraphics{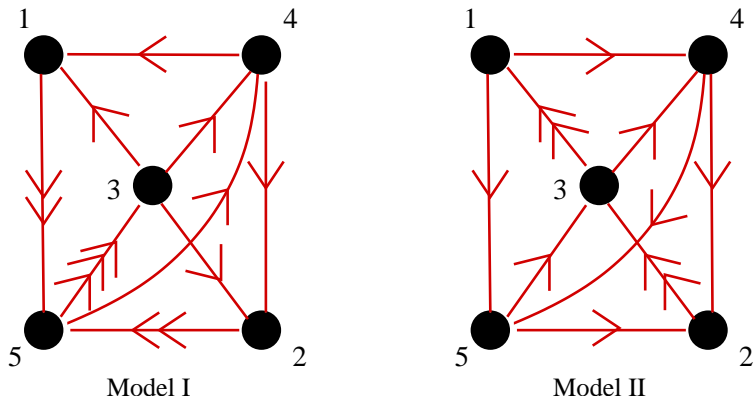}}
\caption{\small The quivers associated to the second del Pezzo surface.}
\label{reg2}
\end{center}
\end{figure}

There are two known phases for the field theory duals, related by Seiberg duality. 
The corresponding quivers are shown in Figure \ref{reg2}.
Both of them have gauge group $SU(N)^5$, but different bi-fundamental field content.

Let us first focus on Model II.\footnote{As in the previous Section we will assume that the nodes of 
the quiver will all be interacting in the IR where the theory will reach a non-trivial fixed point.}
In this model all the bi-fundamental fields are 
flavor singlets but $X^{\alpha}_{31}$ and $X^{\alpha}_{23}$ ($\alpha=1,2$)
which transform in two distinct $SU(2)$'s. The crucial point is that, again, the 
superpotential
\bea
\label{super2}
W&=&\mbox {Tr} \left[X_{34} X_{45}X_{53} - X_{53} X^{2}_{31}X_{15} 
+ X_{34} X_{42}X^2_{23} + \right. \nonumber \\
&& + \left. X^2_{23} X^{1}_{31}X_{15}X_{52} + X_{42} X^1_{23}X^2_{31}X_{14} - 
X^1_{23} X^1_{31}X_{14}X_{45}X_{52}\right]
\eea
breaks this non-abelian flavor symmetry. As a result, the $R$-charges of all the bi-fundamental fields 
are a-priori different.\footnote{As for $dP_1$, in the $a$-maximization
calculations present in the literature for the $dP_2$ dual,
it is always assumed that the $R$-charges of the two doublets are the same, 
i.e. the flavor $SU(2)\times SU(2)$ is unbroken. This leads to the conclusion that 
the $R$-charges and the central charge (and hence the volumes of the dual geometry) are rational.} 
We will now compute such charges and the central charge, as in Section 2. By imposing the vanishing 
of the $\beta$-functions at the IR fixed point and the condition that the superpotential 
should have $R$-charge 2, we get {\it four} independent $R$-charges, in terms of which we 
can write
\bea
\label{sist2}
R_{45} \equiv x \quad,\quad  R_{53}\equiv y &,&  
R_{52}\equiv z \quad,\quad  R^{(2)}_{23}\equiv w \quad,\nonumber \\
R_{14} = w - x \quad,\quad R^{(1)}_{31}=R_{42}&=&x+y-w  \quad,\quad  R^{(2)}_{31}=x+z \quad,\nonumber \\
R^{(1)}_{23}=R_{15}=2-x-y-z &,&
R_{34}=2-x-y\quad.
\eea
Plugging these values in Eq.~(\ref{ace}) we get for the trial $a$-charge
\bea
\label{atrial2}
a_t &=& \frac{9 N^2}{32}\bigl[ 5 + (x-1)^3 +(y-1)^3 +(z-1)^3 +(w-1)^3 + (w -x -1)^3 + \nonumber \\
&&(x+z-1)^3 + 2(x+y-w-1)^3 + (1-x-y)^3 + 2(1-x-y-z)^3
\bigr]
\eea
Extremizing with respect to $x,y,w$ and $z$, we get 
\bea
\label{sol3}
x_{m} = \frac 12 (-5 + \sqrt{33}) &,& y_{m} =  \frac 14 (9 - \sqrt{33})\ , \nonumber \\
w_{m} =  \frac {1}{16} (17 - \sqrt{33}) &,& z_{m} = \frac {3}{16} (19 - 3\sqrt{33}) \ ,
\eea
and hence for the actual central  charge of the theory 
\be
\label{cIR2}
c=a= \frac{243N^2}{1024} \, \left(-59 + 11 \sqrt{33} \right) ~~.
\ee
Note that, as for $dP_1$, the central charge is irrational.
It is larger than the one calculated previously in literature, i.e. $27N^2/28$.
Also it predicts, for the volume of the dual compact space, the value
\be
\label{vol2}
V(X_{dP_2}) = \frac{\pi^3 N^2}{4\,a}= \frac{1}{486} \left(59 + 11 \sqrt{33} \right)\; \pi^3 ~~,
\ee
where we denote as $X_{dP_2}$ the horizon of the complex cone over $dP_2$.\footnote{The above volume is smaller 
than the one for the $dP_1$ case.
For the case of regular $dP_k$ surfaces it is also expected that the related
volumes decrease with $k$ \cite{IW2}.}

The corresponding $R$-charges for the chiral fields can be now easily computed plugging the values (\ref{sol3}) 
in the relations (\ref{sist2}) and are listed in Table 2.
\begin{table} [ht]
\label{tabar2}
\begin{center}
\begin{tabular}{|c|c|}
\hline        
  $X_{ij}$          & $R_{ij}$ \\ 
\hline  
  $X^1_{23}\,,\,X^1_{31}\,,\,X_{42}\,,\,X_{15}$        & $\frac {1}{16} (- 21 + 5\sqrt{33})$ \\  
  $X^2_{23}\,,\,X^2_{31}$        & $\frac {1}{16} (17 - \sqrt{33})$ \\  
  $X_{52}\,,\,X_{14}$          & $\frac {3}{16} (19 - 3\sqrt{33})$ \\  
  $X_{53}\,,\,X_{34}$          & $\frac 14 (9 - \sqrt{33})$ \\    
  $X_{45}$          & $\frac 12 (-5 + \sqrt{33})$ \\  
\hline \end{tabular}
\caption{\small The exact $R$-charges of the bi-fundamental chiral fields for the $dP_2$ dual, Model II.}
\end{center}
\end{table}    
Again, they turn out to be irrational. 
It can be checked that with the above 
$R$-charge assignment there are no violations to the unitarity 
bound.

By using once again the holographic relation (\ref{barvol0}) 
adapted to this case, we can predict the volumes of the supersymmetric three-cycles of the dual geometry. 
Using Eq.~(\ref{vol2}) and Table 2, these turn out to be, again, irrational.

As far as Model I is concerned, this is Seiberg dual to Model II, so the physical quantities in 
the IR such as the central charge should be the same. One can repeat the calculations above 
and check that this is indeed the case. This is a non-trivial check, since the two theories look 
pretty different. As expected, the equality holds for $a_t(max)$. 
Finally, the $R$-charges of the chiral fields are again irrational and are listed in Table 3.
\begin{table} [ht]
\label{tabar3}
\begin{center}
\begin{tabular}{|c|c|}
\hline        
  $X_{ij}$          & $R_{ij}$ \\ 
\hline  
$X_{32}\,,\,X^2_{15}\,,\,X_{31}\,,\,X^2_{25}$ & $\frac{1}{16} (-21 + 5\sqrt{33})$ \\  
$X^1_{53}\,,\,X^2_{53}$          & $\frac 14 (9 - \sqrt{33})$ \\  
  $X_{41}\,,\,X_{42}$            & $\frac {3}{16} (19 - 3\sqrt{33})$ \\  
  $X^1_{15}\,,\,X^1_{25}$        & $\frac {1}{16} (17 - \sqrt{33})$ \\  
  $X_{34}$                       & $\frac {1}{2} (- 5 + \sqrt{33})$ \\ 
  $X_{54}$                       & $\frac {1}{8} (- 21 + 5\sqrt{33})$ \\
  $X^3_{53}$                       & $\frac {1}{8} (37 -5 \sqrt{33})$ \\ 
 \hline \end{tabular}
\caption{\small The exact $R$-charges of the bi-fundamental chiral fields for the $dP_2$ dual, Model I.}
\end{center}
\end{table}    

Let us end with a comment. 
The value of the volume of $X_{dP_2}$, Eq.~(\ref{vol2}), is close to the one of $Y^{3,1}$. 
From the general formula (\ref{volume}) this reads
\be
\label{vol3}
V(Y^{3,1}) = \frac{1}{648} \left(63 + 11 \sqrt{33} \right)\; \pi^3 ~~.
\ee
This is smaller than $V(X_{dP_2})$, and thus $ a(X_{dP_2}) < a(Y^{3,1})$. The SCFT dual to 
$Y^{3,1}$ has $SU(N)^6$ gauge group 
\cite{MS}, while for the dual of $X_{dP_2}$ this is $SU(N)^5$. 
What one can argue is that the two SCFT's may be related by some higgsing. It would be clearly
interesting to find an explicit realization of this higgsing and a metric for $X_{dP_2}$.

\vskip 15pt
\centerline{\bf Acknowledgments}
\vskip 10pt
\noindent
We thank B. Acharya, D. Martelli, J. Sparks and Y. Takanishi 
for discussions and exchange of ideas. We are grateful to K. Intriligator for 
very interesting remarks and e-mail correspondence and to A. Schwimmer for 
useful comments. We finally thank the authors of \cite{MS} for communicating their 
geometric findings prior to publication, and for earlier collaboration on related material. 
Work partially  supported
by the European Commission RTN Program MRTN-CT-2004-005104, MRTN-CT-2004-503369,
EC Excellence Grant MEXT-CT-2003-509661
and by MIUR. M.B. is also supported by a MIUR fellowship within the program 
``Incentivazione alla mobilit\`a di studiosi stranieri e italiani residenti all'estero''.

\end{document}